\documentclass[aps,preprint,showpacs,superscriptaddress,groupedaddress,amsmath]{revtex4}

\usepackage{amsmath,amsfonts,graphicx,bm,amssymb,array}
\usepackage[usenames]{color}
\usepackage{array}
\usepackage{float}
\usepackage{sidecap}
\usepackage{graphicx}
\usepackage{bm}
\usepackage{axodraw4j}
\usepackage{color}
\usepackage[section]{placeins}
\usepackage{booktabs}

\def\etmiss{E\!\!\!\!\slash_{T}}

\def\pslash{\not{\hbox{\kern-4pt $p$}}}
\def\qslash{\not{\hbox{\kern-4pt $q$}}}
\def\lv{\not{\hbox{\kern-4pt $L$}}}
\def\lsim{\mathrel{\raise.3ex\hbox{$<$\kern-.75em\lower1ex\hbox{$\sim$}}}}
\def\gsim{\mathrel{\raise.3ex\hbox{$>$\kern-.75em\lower1ex\hbox{$\sim$}}}}
\def\ifmath#1{\relax\ifmmode #1\else $#1$\fi}

\definecolor{DarkRed}{rgb}{0.55,0.00,0.00}

\newcommand{\nc}{\newcommand}
\nc{\postscript}[2]{\setlength{\epsfxsize}{#2\hsize}\centerline{\epsfbox{#1}}}
\nc{\beq}{\begin{equation}}   \nc{\eeq}{\end{equation}}
\nc{\bea}{\begin{eqnarray}}   \nc{\eea}{\end{eqnarray}}
\nc{\baa}{\begin{array}}      \nc{\eaa}{\end{array}}
\nc{\bit}{\begin{itemize}}    \nc{\eit}{\end{itemize}}
\nc{\ben}{\begin{enumerate}}  \nc{\een}{\end{enumerate}}
\nc{\bce}{\begin{center}}     \nc{\ece}{\end{center}}
\nc{\non}{\nonumber}
\nc{\eqns}[1]{\begin{eqnarray} #1 \end{eqnarray}}
\nc{\braket}[1]{\left( #1 \right)}

\begin{document}

\preprint{MSUHEP-120723}
\title{Single Top Production as a Probe of B$'$ Quarks}

\author{Joseph Nutter} \affiliation{Department of Physics and Astronomy, Michigan State University, East Lansing MI 48824, USA}
\author{Reinhard Schwienhorst} \affiliation{Department of Physics and Astronomy, Michigan State University, East Lansing MI 48824, USA}
\author{Devin G.\! E.\! Walker} \affiliation{SLAC National Accelerator Laboratory, 2575 Sand Hill Road, Menlo Park, CA 94025, USA}
\author{Jiang-Hao Yu} \affiliation{Department of Physics and Astronomy, Michigan State University, East Lansing MI 48824, USA}

\date{\today}

\begin{abstract}
We show how single top production at the LHC can be used to discover (and characterize the 
couplings of) $B'$~quarks, which are an essential part of many natural models of new 
physics beyond the Standard Model.  We present the $B'$~effective model and concentrate on 
resonant production via a colored anomalous magnetic moment.  Generally, $B'$s preferentially 
decay into a single top quark produced in association with a $W$~boson; thus, this production 
process makes associated single top production essential to $B'$~searches at the LHC.  
We demonstrate the background processes are manageable and the signal cross section 
is sufficient to yield a large signal significance even during the 7~TeV LHC run.  Specifically, 
we show that $B'$~masses of 700~GeV or more can be probed.  Moreover, if a $B'$ is found, then 
the chirality of its coupling can be determined.  Finally, we present signal cross sections 
for several different LHC energies.
\end{abstract}

%activate the following line for publication
\maketitle

% \modulolinenumbers[1]
% \linenumbers

%%%%%%%%%%
%%%%%%%%%%    Main Text
%%%%%%%%%%

\section{Introduction}
\label{sec:intro}

For the first time in history, the TeV scale is being directly probed by the Large Hadron 
Collider (LHC).  Remarkably, a 125~GeV Higgs-like 
particle~\cite{atlashiggsresults,CMShiggsresults} has already been discovered.  Beyond this  
tremendous achievement, an additional focus of the LHC is to uncover new particles which 
will presage new physics scenarios. Fourth generation $B'$~quarks are an example of such a 
new particle. They are essential to many new physics scenarios and could appear
as a chiral or vector-like quark.  
A sample of popular models with $B'$~quarks (or the strongly coupled equivalent) can be found 
in~\cite{technicolor,topcolor,topseesaw,tcreview,Kaplan:1983sm,LH,Arkani-Hamed:1998rs,Randall:1999ee}.  
In Section~\ref{sec:intersection}, we further detail the role $B'$~quarks in natural models
 and focus on their outsized importance in model building in avoidance of precision 
electroweak constraints.  

Final states involving single top quarks provide an important discovery mode for 
$B'$~quarks that has not been explored heretofore. Moreover, given sufficient data, the single 
top final state is uniquely sensitive to the chirality of the $B'$~quark.
The LHC has been remarkably adept at searching for single top events.  About three million 
single top events should have been recorded (after cuts) during the 7~TeV LHC 
run~\cite{Chatrchyan:2011vp,Aad:2012ux}.  Additionally, evidence for SM $W\,t$ production was 
presented~\cite{ATLAS:2012dj}. More importantly, the production cross section for SM single 
top events is precisely known at NLO with NNLO corrections~\cite{Kidonakis:2011wy}.   
We present an effective model for $B'$ quark production and decay during the LHC runs at 
7~TeV, 8~TeV and 14~TeV center-of-mass (c.m.) energies. We explore the signal
and backgrounds for a $B'\rightarrow W\,t$ single top search in the lepton+jets final state.
This work should be considered as complementary to $B'$ pair production which subsequently 
decays to top quark pairs.  The combination of the two signatures should be part of a 
comprehensive plan to maximizes the sensitivity of the LHC to natural new physics.  To date, 
the $B'$ searches from the Tevatron and the LHC have relied exclusively on the pair-production 
mode, in searches for SM-like decays of the fourth generation
quark~\cite{ATLAS:2012aw,Chatrchyan:2012yea,Aaltonen:2011vr} and in
searches for chiral and vector-like $B'$ quarks~\cite{ATLAS:2012ak,Aad:2011yn,Abazov:2010ku}.

This paper is organized in the following way: In Section~\ref{sec:intersection}, we describe 
the constraints and implication of $B'$~models on new physics beyond the Standard Model.
Next in Section~\ref{sec:models}, we outline our effective $B'$~models and conventions.  
We detail a benchmark scenario which is simulated and analyzed in the subsequent sections. 
Section~\ref{sec:xsec} gives the cross section for the $p\,p\rightarrow B'\rightarrow W\,t$ 
process. The phenomenological analysis is described in Section~\ref{sec:analysis}, including
the various backgrounds expected at the LHC. Finally we conclude.

\section{B' Models, Naturalness and Precision Measurements}
\label{sec:intersection}

\subsection{Naturalness and New Physics Scales}

It is well known that a light 125~GeV Higgs boson~\cite{atlashiggsresults,CMShiggsresults} 
illustrates a serious theoretical inconsistency in the Standard Model.  Radiative corrections, 
generated dominantly by top quark loops, push the Higgs boson to have a mass of order the next 
largest scale of new physics. Thus, since the Planck scale is only known scale beyond the 
weak scale, naively the Higgs boson should have a mass of order $10^{19}$ GeV.  This implies 
the couplings in the Higgs potential must be severely fine-tuned in order to get the right 
electroweak symmetry breaking vev.  Natural models of new physics solve this problem by adding 
new top partners to the SM which cancel (some or all of) the top quark radiative corrections.  
These top partners \textit{must} be in an electroweak doublet in order to properly cancel the 
divergences to the Higgs mass by the SM top quark. Thus, many natural models also feature 
$B'$~quarks, the heavy partner of the bottom quark. Consequently, discovering a $B'$~quark may 
be a harbinger of new natural physics beyond the SM.  Moreover, if the top and bottom partners 
have the same mass hierarchy as the SM top and bottom, the $B'$~quark may be the first to be 
discovered.  

There is more to this story of top partners, naturalness and new heavy quarks. It has been 
shown~\cite{Chanowitz:1978uj,Chanowitz:1978mv} that partial wave unitarity can place an upper 
bound on the mass of additional heavy fermions which obtain \textit{all} of their mass from 
electroweak symmetry breaking.  For heavy $B'$~quarks, this limit is
\begin{equation}
m_{B'} < 500/\sqrt{N} \,\,\mathrm{GeV}
\end{equation}
where $N$ counts the number of degenerate $SU(2)_L$ doublets.  Natural models get around this 
bound by requiring a new scale of physics~\cite{technicolor,topcolor,topseesaw,tcreview,Kaplan:1983sm,LH,Arkani-Hamed:1998rs,Randall:1999ee}. We show that 500~GeV $B'$~quark masses can easily 
be seen during the 7~TeV run at the LHC.  A heavier $B'$~quark would imply a new scale of 
physics beyond the SM.  It also implies that the 7~TeV LHC run can rule out traditional fourth 
generation $B'$s which get all of their mass from electroweak symmetry breaking.

\subsection{Bounds on B' Models from Precision Measurements}
~
Models with exact custodial symmetry generate minimal corrections to the well-constrained 
$S$ and $T$ parameters~\cite{Peskin:1991sw}.  Custodial symmetries are therefore a common 
feature of natural new physics beyond the Standard Model.  Because of this, and the fact that 
natural models feature a large coupling between the top partner and the SM, implicitly the 
bottom partner also has a significant coupling~\footnote{In the limit of exact custodial 
symmetry, the mass of the top and bottom partners are the same.  Thus, the $T$ parameter goes 
to zero.}.  After spontaneous symmetry breaking, the top and bottom partners 
can mix with the SM third generation.  This mixing can potentially lead to large corrections
to $Z \to \overline{b}\, b$. Indeed, precision measurements from LEP and SLAC require 
less than $0.3\%$ deviation~\cite{Nakamura:2010zzi} from the SM prediction for this process.
Yet new models of electroweak symmetry breaking (e.g. extra-dimensional scenarios) can 
generate $20$-$40\%$ corrections. It was recognized that an 
``extended'' custodial symmetry could be arranged to prevent large corrections to 
$Z \to \overline{b}_L b_L$ but not $Z \to \overline{t}_L \,t_L$ and $W \to \overline{t}_L \,b_L$
simultaneously~\cite{Agashe:2006at}.  This symmetry is
\begin{equation}
O(4) \sim SU(2)_L \times SU(2)_R \times P_{LR},
\end{equation}
where $P_{LR}$ is a parity interchanging left and right.  It has also been shown there can 
be tensions between $Z \to \overline{b}\, b$ constraints and the $T$ parameter for another 
extended custodial symmetry~\cite{SekharChivukula:2009if}. Our single top signal directly 
probes the $W \to \overline{t}_L \,b_L$ coupling and the mixing between the $B'$~and 
$b$~quarks.  Thus, the nature of the custodial symmetry is a consequence of the search for 
bottom partners.

\section{Effective Couplings and Conventions }
\label{sec:models}

\subsection{Effective Lagrangian}
\label{sec:lag}
~
To probe $B'$~models, we consider an effective scenario where a new $B'$~quark is the only 
light state below a cutoff~\footnote{This scenario is possible if, above the cutoff, 
new quarks transforming as a $N$ of $SU(2)_L$ with hypercharges $Y = \frac{1}{6}(1 - 3 N), \,\, \frac{1}{6}(1 - 3 (N - 2)),\,\,  \frac{1}{6}(1 - 3 (N - 4)), \ldots$, are added in various 
anomaly free combinations.  Here the number of different hypercharges is the dimension of $N$. 
After (all of) the symmetry breaking, the fermions with $Q = -1/3$ mixes with the SM bottom 
quark. }.  The most general Lagrangian describing the interactions of heavy bottom quarks with 
gluons (assuming operators of dimension five or less) is~\cite{Baur:1987ga}
\begin{equation}
	{\mathcal L} =  g_s\,\overline{B'} \gamma^\mu G_\mu B' + \frac{g_s\,\lambda}{2\,\Lambda} G_{\mu\nu}\, \overline{b}\, \sigma^{\mu\nu} \biggl(\kappa^b_L P_L + \kappa^b_R P_R\biggr) B' + \mathrm{h.c.} \label{eq:productionL}
	\end{equation}
The dimension five operator is generated in many models~\cite{technicolor,topcolor,topseesaw,tcreview,Kaplan:1983sm,LH,Arkani-Hamed:1998rs,Randall:1999ee} by integrating out new states.
Here we follow convention and set the scale $\Lambda$ to $M_{B'}$.  $P_{L,R}$ are the normal 
projection operators, while $\lambda$ is a free parameter whose value is dependent on the 
UV physics that was integrated out.  We focus on the coupling with gluons because of the 
large fraction of  gluon initial state partons at LHC energies.  Similar operators
can generate flavor-changing-neutral-currents (FCNC).  We assume the UV theory is 
free of FCNCs, therefore ensuring $\lambda$ is sufficiently suppressed.

The electroweak decay of the $B'$~quark into a single top quark can be parametrized as
\begin{equation}
	{\mathcal L} = \frac{g_2}{\sqrt{2}} \,W^+_{\mu} \,\overline{t} \gamma^{\mu}\, \bigl(f_L P_L + f_R P_R \bigr) \,B' + \mathrm{h.c.}
\end{equation}
Here $g_2$ is the $SU(2)_L$ coupling.  In the case where a left-handed (chiral or vector-like) 
quark mixes with the left-handed bottom quark, the couplings are
\begin{equation}
	f_L = s_L, \quad f_R \simeq 0\; .
\end{equation}
For the right-handed case, the couplings are
\begin{equation}
	f_L \simeq 0, \quad f_R = s_R\; .
\end{equation}
The partial decay width of the $B'$~quark to $Wt$ is given by
\begin{equation}
	\Gamma (B \to t\, W^- ) = \frac{g_2^2}{64\pi} \frac{M_B^3}{M_W^2} (f_L^2 + f_R^2) (1  -  x_t^2)^3 + {\mathcal O}(x_W^2) \; .
\end{equation}

We consider a $B'$ benchmark scenario with couplings 
$s_L = s_R= \frac{v}{m_{B'}}$, $\kappa_L = \kappa_R = 0.5$. With these settings, the total 
decay width at a $B'$~mass of 700~GeV is 31.85~GeV.
The branching ratio, together with those for $bg$, $bZ$ and $bH$ decays are shown in 
Fig.~\ref{fig:br}. (See the Appendix for those partial decay widths.)  At low masses, 
the $bZ$ and $bH$ decays dominate, while at higher masses
the $Wt$ decay is the largest and approaches 40\% of the total width. The large decay
branching ratio to $Wt$ makes this an attractive final state for a $B'$~search.

\begin{figure}[!tbp]
\begin{center}
	\includegraphics[scale=0.6]{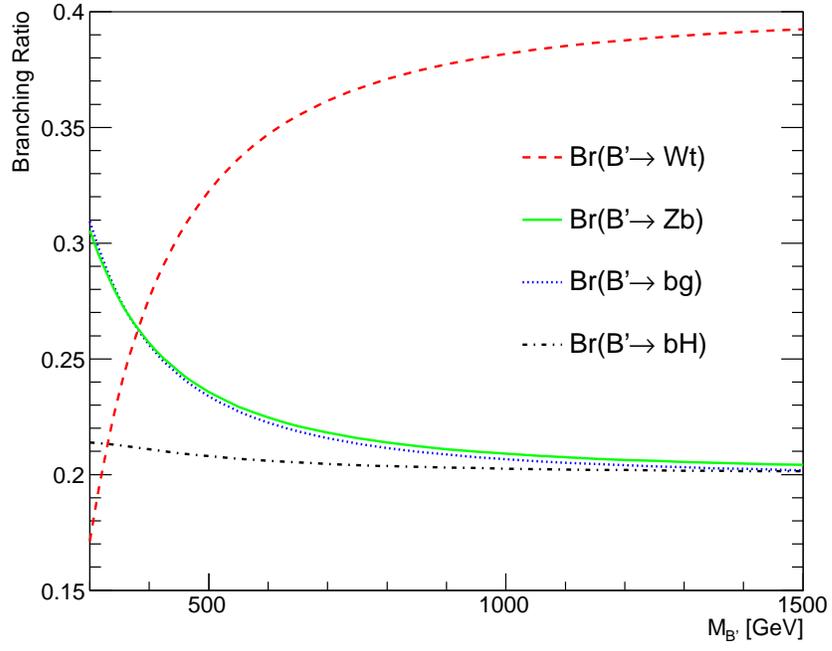}
	\caption{The decay branching ratios of $B'$ to $Wt$, $bg$, $bZ$ and $bH$, as a
function of the $B'$~mass. The couplings are given in the text.}
	\label{fig:br}
\end{center}
\end{figure}

%%%%%%%%%%%%%%%%%%%%%%%%%%%%%%%%%%%%%%%%%%%%%%%%%%%%%%%%%%%%%%%%%%%%%%%%%%%%%%%%%%%%%%%%%%%%%%%%%%%
%   Section on the production cross section and its dependence on various parameters
%%%%%%%%%%%%%%%%%%%%%%%%%%%%%%%%%%%%%%%%%%%%%%%%%%%%%%%%%%%%%%%%%%%%%%%%%%%%%%%%%%%%%%%%%%%%%%%%%%%
\clearpage

\section{$B'$ Production and Decay to $tW$}
\label{sec:xsec}
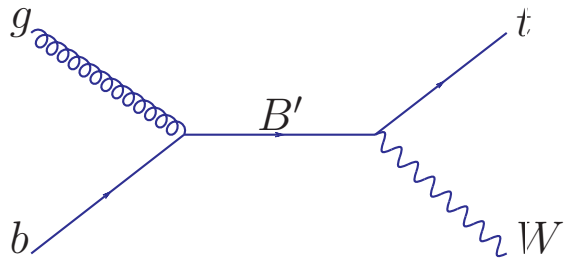
\begin{figure}[!tbp]
\begin{center}
\fcolorbox{white}{white}{
\SetScale{0.4}\setlength{\unitlength}{0.4pt}
  \begin{picture}(452,300) (95,-100)
    \SetWidth{2.0}
    \SetColor{Blue}
    \Gluon(96,128)(240,32){7.5}{13}
    \Line[arrow,arrowpos=0.5,arrowlength=8,arrowwidth=2,arrowinset=0.8](96,-80)(240,32)
	\Line[arrow,arrowpos=0.5,arrowlength=8,arrowwidth=2,arrowinset=0.8](240,32)(420,32)
    \Line[arrow,arrowpos=0.5,arrowlength=8,arrowwidth=2,arrowinset=0.8](420,32)(544,128)
	\Photon(420,32)(544,-80){7.5}{8}
	 \Text(76,125)[lb]{\Large{\Black{$g$}}}
     \Text(76,-80)[lb]{\Large{\Black{$b$}}}
	 \Text(310,36)[lb]{\Large{\Black{$B'$}}}
    \Text(555,125)[lb]{\Large{\Black{$t$}}}
    \Text(555,-80)[lb]{\Large{\Black{$W$}}}
   %\Text(310,-135)[lb]{\Large{\Black{$(a)$}}}
  \end{picture}
}
\end{center}
\caption[]{Feynman diagrams for production of a fourth generation $B'$ quark and decay to
a top quark and $W$~boson.}
\label{fig:stop_bprime}
\end{figure}
~
We consider the production of $B'$~quarks via the following process,
\begin{equation}
p + p \to B'  \to t + W. 
\end{equation}
The Feynman diagram for resonance $B'$~production and decay to $Wt$ is shown in 
Fig.~\ref{fig:stop_bprime}. 
This process relies on $b$~quark and gluon initial partons.  The gluon parton helps the 
cross section tremendously; however, this process is suppressed by a dimension five anomalous 
magnetic moment operator (see equation~\ref{eq:productionL}) and the $b$~quark initial partons.
Notably, the $B'$~quark has access to the full center-of-mass energy of the colliding partons.
This increases the ability to probe heavy $B'$~quarks in contrast to $B'$~quark
pair production.  

\begin{figure}[!tbp]
\begin{center}
	\includegraphics[scale=0.7]{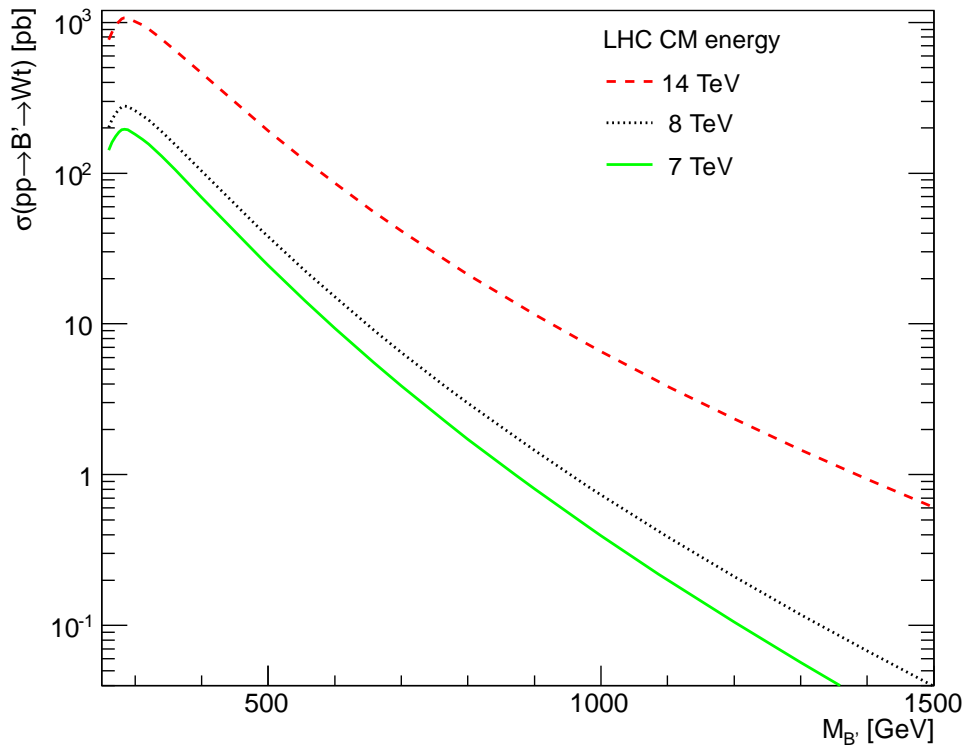}
	\caption{The total cross section for $B'$ production with decay to $tW$ at the
          LHC, for $f_L=1$, $f_R=0$ and three different c.m. energies.}
	\label{fig:xsec}
\end{center}
\end{figure}

In this section we explore the $t\,W$ final state. We also present the production cross 
section for different $B'$~masses at the LHC for three different beam energies.  The total 
cross section for $B'$~production and subsequent decay to a top quark and $W$~boson
at the LHC are computed with Madgraph~\cite{Alwall:2011uj} for several different c.m. energies
and the same couplings as in Sec.~\ref{sec:lag},
and are shown in Fig.~\ref{fig:xsec}.  We use the CTEQ6L1 set of parton distribution 
functions (PDF)~\cite{Pumplin:2002vw} and set the 
factorization and renormalization scales to the $B'$~mass.
The cross section peaks at about 300~GeV, where the top quark and $W$~boson are both on 
shell, and then decreases
for higher $B'$~masses. This decrease is due in part to the decrease of the parton luminosity
and in part to the decreasing $B'gb$ coupling as the $B'$ mass increases. 
The increasing parton luminosity
is visible also in different slopes for the three curves, more so when comparing 14~TeV
to the other two. The uncertainty on the cross section for a $B'$~mass of 700~GeV is
0.3\% when varying the top quark mass by 1~GeV and 13\% when varying the factorization and 
renormalization scales up and down by a factor of two. 
% The PDF uncertainty is ???\%, using the set of CTEQ uncertainty eigenvectors. 
~

%%%%%%%%%%%%%%%%%%%%%%%%%%%%%%%%%%%%%%%%%%%%%%%%%%%%%%%%%%%%%%%%%%%%%%%%%%%%%%%%%%%%%%%%%%%%
%   signals and backgrounds in a benchmark scenario
%%%%%%%%%%%%%%%%%%%%%%%%%%%%%%%%%%%%%%%%%%%%%%%%%%%%%%%%%%%%%%%%%%%%%%%%%%%%%%%%%%%%%%%%%%%%
%
\section{Analysis}
\label{sec:analysis}

The LHC has collected enough events already to look for singly produced $B'$. Here, an example 
analysis demonstrates the prospects for observing a $B'$~quark at the 7~TeV LHC. We consider
the lepton+jets $B'$ final state and evaluate the backgrounds to this signature. We look at 
the process $p p \to B \to t \bar{b} \to b l^+ \nu \bar{b}$. $B'$~signal events are produced
at a benchmark mass of 700~GeV and the numbers of signal and background events
remaining after basic selection cuts are computed.

Signal and background events are generated with Madgraph~\cite{Alwall:2011uj} and are
normalized to the corresponding LO cross sections. The dominant
backgrounds to the final state of lepton and three jets are from top quark pair production and
$W$~boson production in association with jets. For top pair production we include both the 
lepton+jets final state, $t\bar{t} \to  b l \nu\, \bar{b} jj$, and the dilepton+jet final 
state, $t\bar{t}j \to  b l \nu\, \bar{b} l \nu\, j$. For the lepton+jets final state, one of
the jets must be at low $P_T$ or otherwise be lost in order to enter the signal region. 
For the dilepton+jet final state, one of the leptons must be at low $P_T$ or otherwise
be lost. Smaller backgrounds are from single top quark production in association with a 
$W$~boson ($t+W$) or with jets ($t+jets$, $t$-channel and $s$-channel) and from 
diboson+jet ($WV$) production. 

We use the anti-kt algorithm in the FastJet~\cite{Cacciari:2011ma} package to cluster quarks
and gluons into final state jets with parameter $R = 0.4$. 
Detector resolution effects are simulated by smearing jet and leptonic energies according to a 
Gaussian:
\begin{eqnarray} 
\frac{\delta E}{E}= \frac{\mathcal{A}}{\sqrt{E/{\rm GeV}}}\oplus \mathcal{B}\,,
\label{eqDQ:smear}
\end{eqnarray}
where $\frac{\delta E}{E}$ is the energy resolution, $\mathcal{A}$ is a sampling term, 
$\mathcal{B}$ is a constant term, $\oplus$ represents addition in quadrature, and all 
energies are measured in GeV.  For leptons we take $\mathcal{A}=5\%$ and 
$\mathcal{B}=0.55\%$, while for jets we take $\mathcal{A}=100\%$ and 
$\mathcal{B}=5\%$, chosen to represent the ATLAS and CMS detector 
performance~\cite{Aad:2008zzm,Ball:2007zza}.
We do not smear $\etmiss$.
We model $b$-tagging as a flat 60\% probability to tag $b$-quark jets and a 0.5\% probability
to mistag non-$b$-quark jets (including charm quarks).

Signal and background events are required to pass the following basic selection cuts:
\begin{eqnarray}
\textrm{At least two jets with } \qquad p_T^j&\geq& 25\,{\rm GeV}, \qquad 
\left|\eta_{j}\right|\leq 2.5 \nonumber \\
\textrm{Exactly one lepton with } \qquad p_{T}^{\ell}&\geq& 25\,{\rm GeV}, \qquad
\left|\eta_{\ell}\right|\leq 2.5, \nonumber \\
\textrm{Missing energy } \qquad  ~\etmiss &>& 25~{\rm GeV},  \nonumber \\
\textrm{Object separation } \qquad \Delta R_{jj,j\ell} &>& 0.4, 
\qquad \qquad \Delta R_{\ell\ell} > 0.2.
\label{eqDQ:basiccut}
\end{eqnarray}

\begin{figure}[!h!tbp]
\begin{center}
	\includegraphics[scale=0.4]{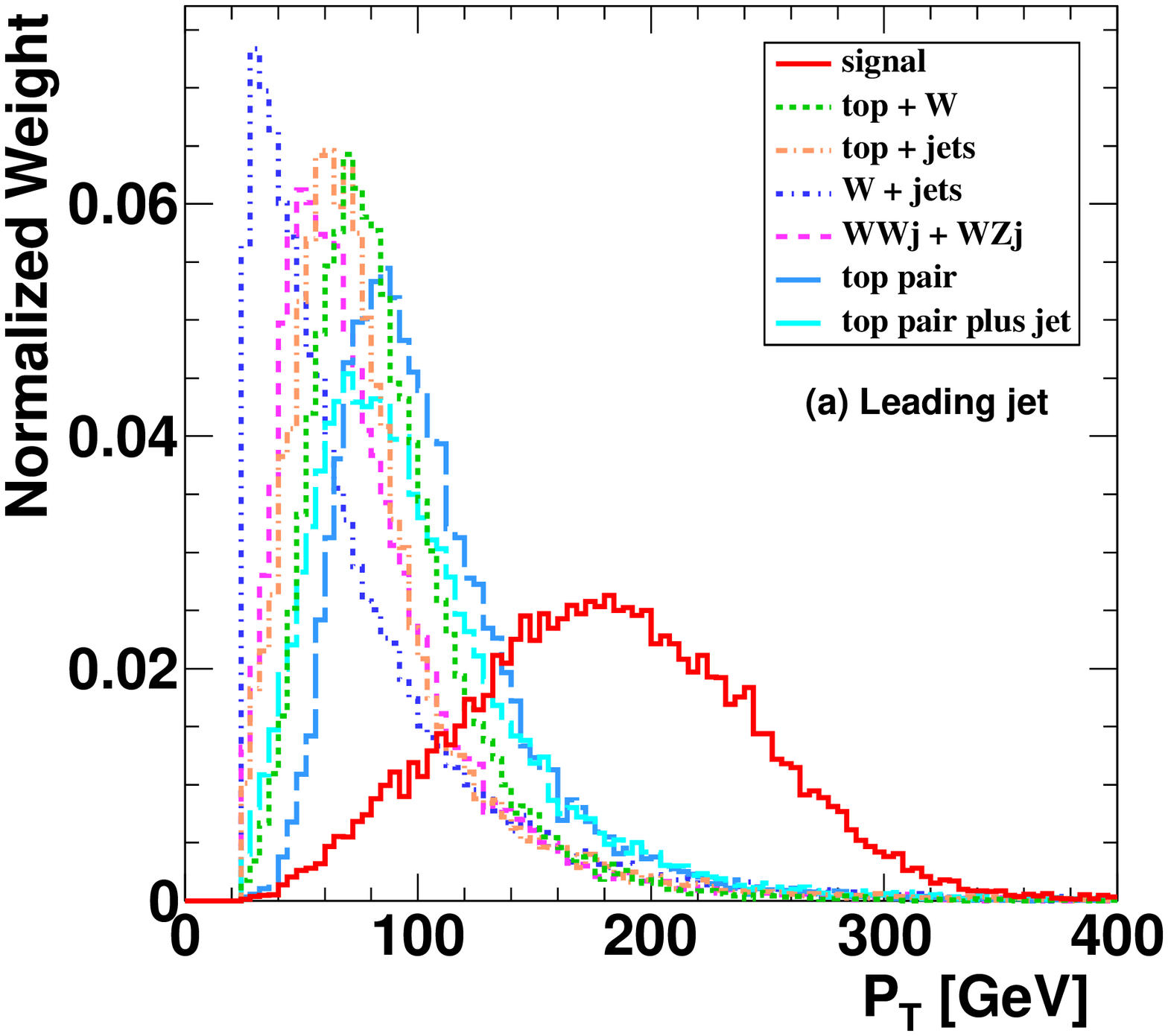}
	\includegraphics[scale=0.4]{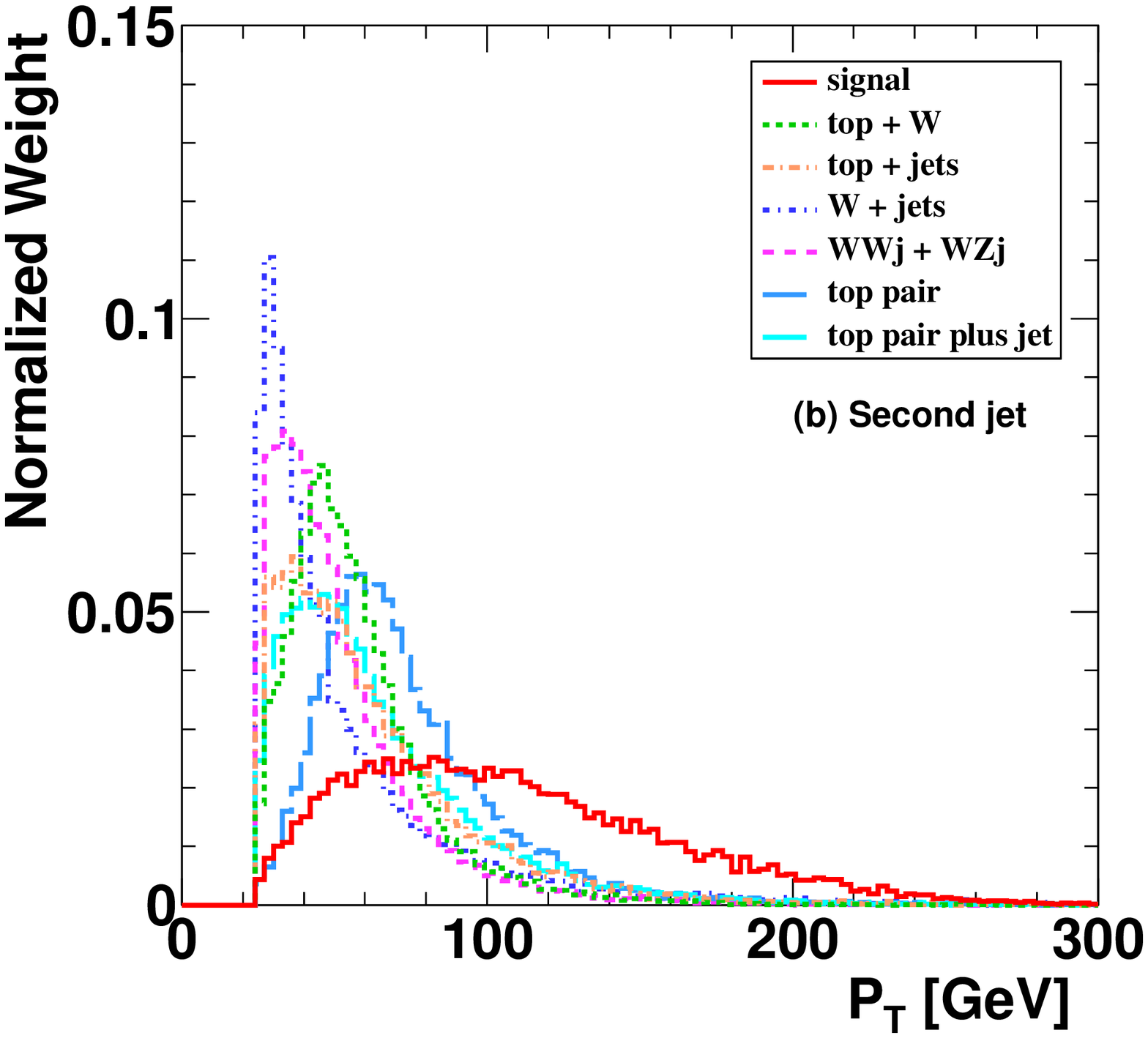}
	\includegraphics[scale=0.4]{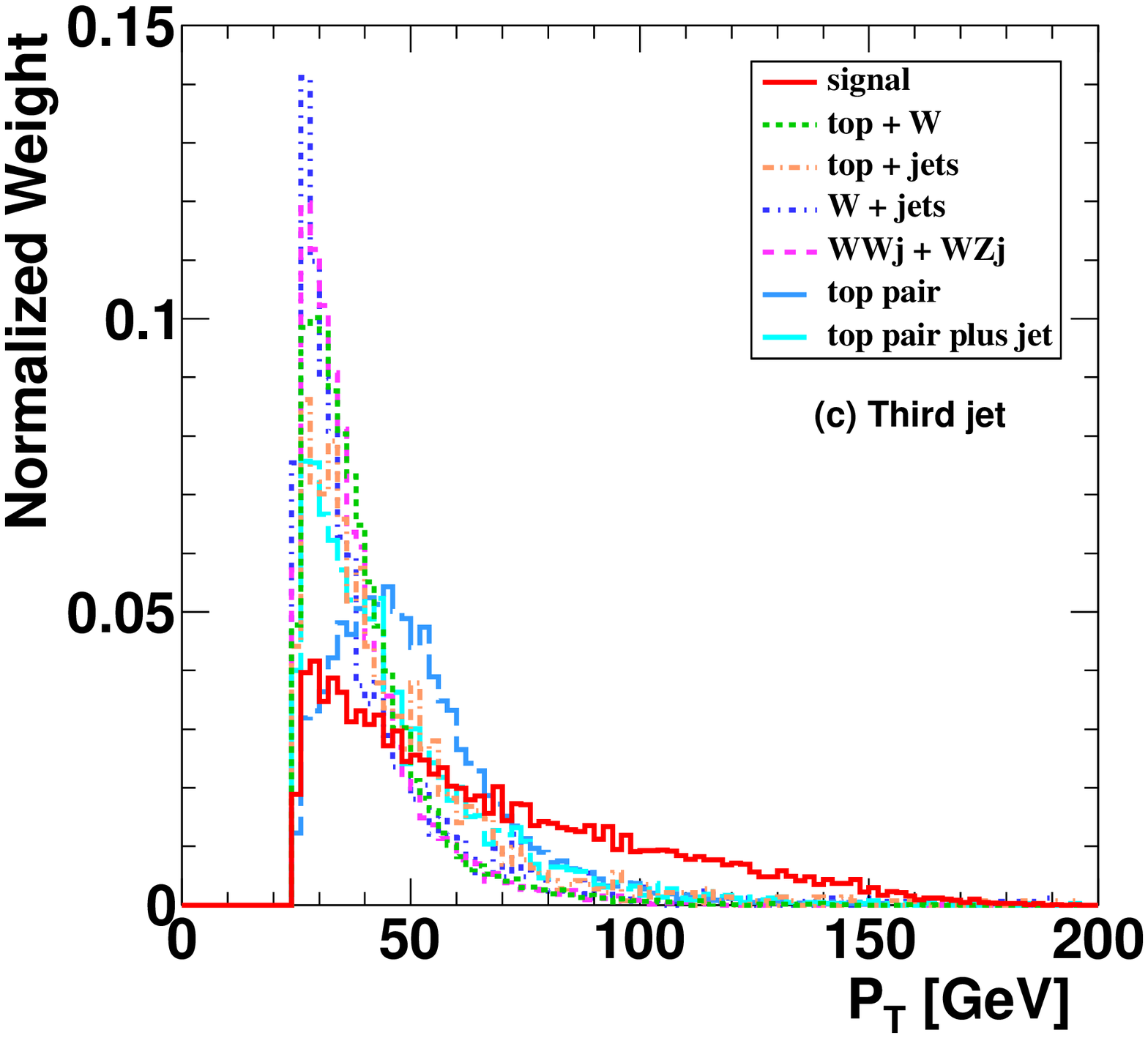}
	\includegraphics[scale=0.4]{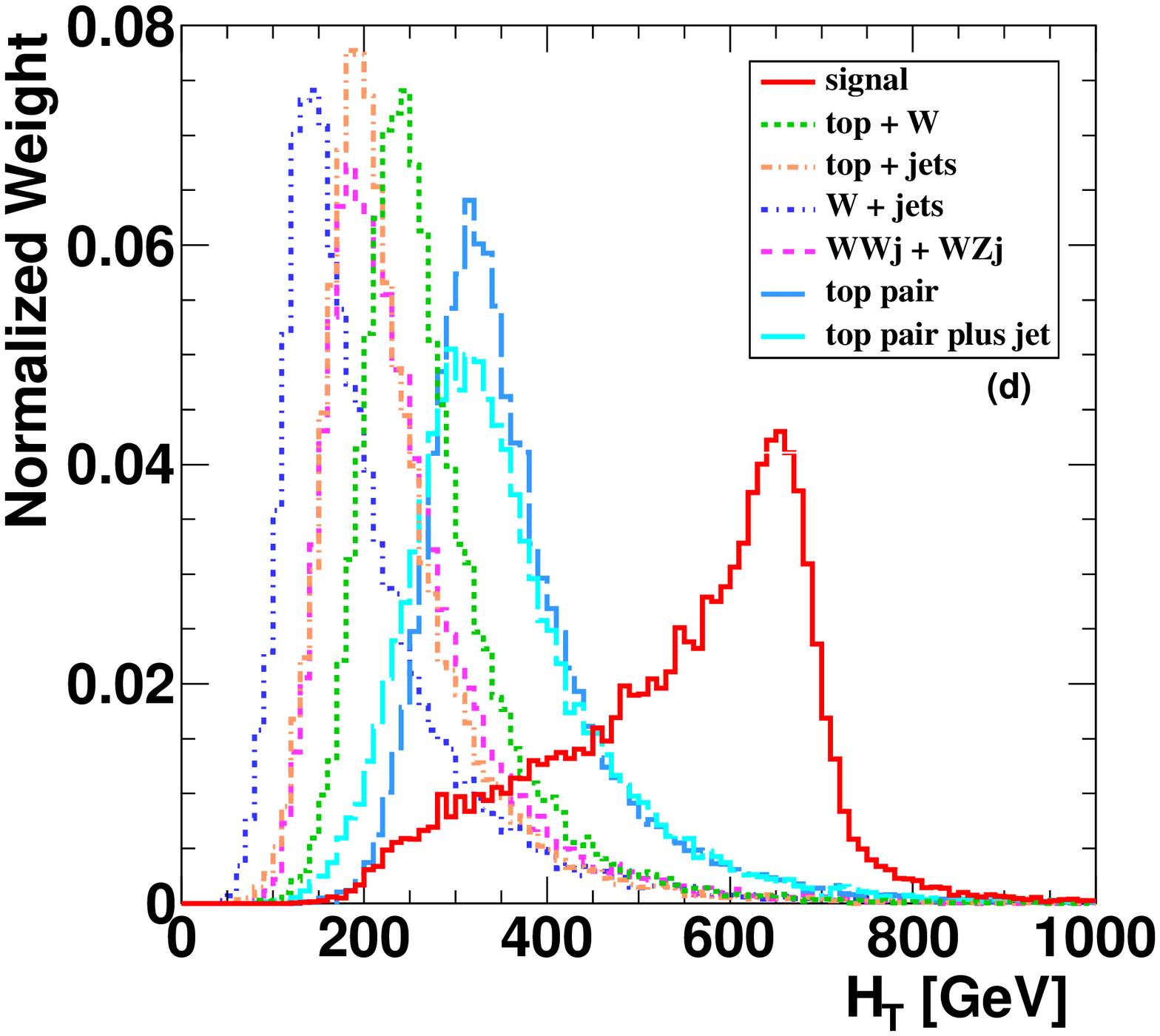}
	\caption{The distribution of (a) $P_{T}$ of the leading jet, (b) $P_{T}$ of the second
jet, (c) $P_{T}$ of the third jet and (d) $H_T$ of all final state objects 
for the $B'$ signal and backgrounds after basic selection cuts.
Each distribution is normalized to unit area.}
	\label{fig:jetpt}
\end{center}
\end{figure}

The kinematic distribution of the $B'$ signal and the various backgrounds after these
cuts are shown in Fig.~\ref{fig:jetpt}. The backgrounds are mostly
at low $P_T$, whereas the $B'$ signal is at high $P_T$. The top quark pair background 
extends farthest into the $B'$ signal region. This can also be clearly seen in the distribution
of $H_T$, the scalar sum of the $P_T$ of all final state objects.
~

To isolate the $B'$ signal and suppress the SM backgrounds, a set of final cuts is applied on
the jet $P_T$ and on $H_T$,
\begin{eqnarray}
p_T^{\rm jet\, 1}&\geq& 80\,{\rm GeV},  \nonumber \\
p_T^{\rm jet\, 2}&\geq& 50\,{\rm GeV},  \nonumber \\
p_T^{\rm jet\, 3}&\geq& 40\,{\rm GeV},  \nonumber \\
H_T              &\geq& 425\,{\rm GeV}.
\label{eqDQ:ptcut}
\end{eqnarray}
To suppress the background from $W$+jets and dibosons further, we require at least one 
jet to be $b$-tagged. 

\begin{figure}[!h!tbp]
\begin{center}
	\includegraphics[scale=0.68]{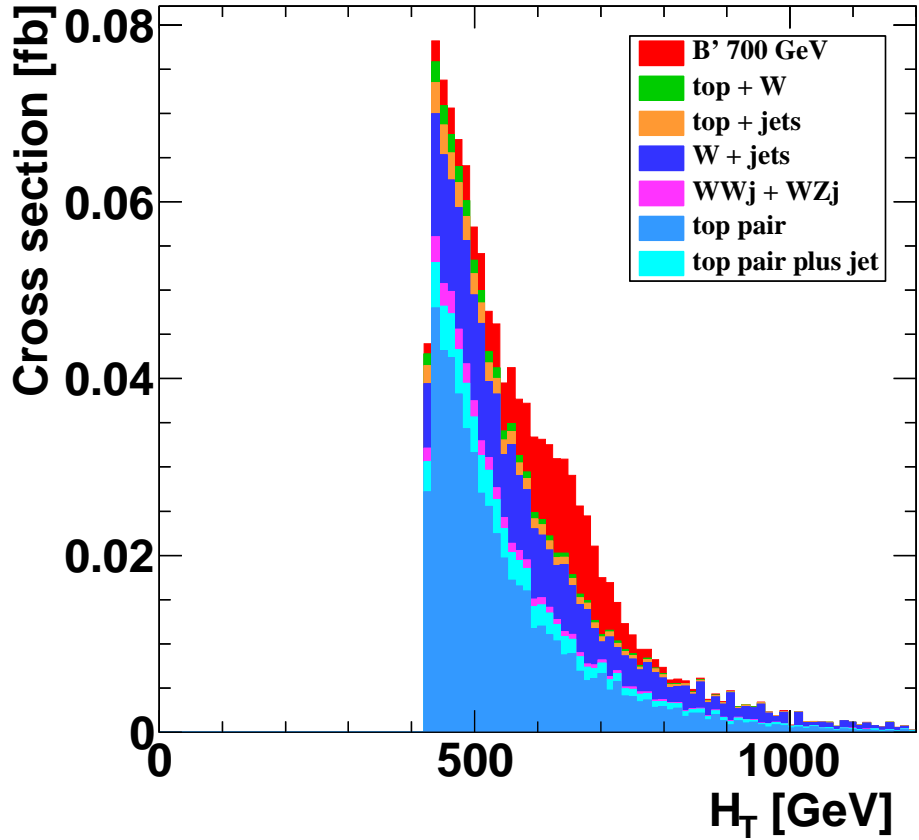}
	\caption{$H_T$ distribution for the $B'$ signal and backgrounds normalized to
        their production cross sections after the final cuts.}
	\label{fig:sysht}
\end{center}
\end{figure}

These cuts effectively suppress most of the SM backgrounds while passing much of the 
$B'$ signal. 
The distribution of $H_T$ after the final cuts is shown in Fig.~\ref{fig:sysht}.
The largest remaining background contribution is from top pair production. At low $H_T$, 
$W$+jets also contributes, less so at high $H_T$. 
~

The final set of cuts effectively isolates a $B'$~signal at any mass above 600~GeV.
In order to further improve the sensitivity of the analysis, the reconstruction of the 
$B'$~quark and its invariant mass is required. For this reconstruction it is necessary to 
first obtain the neutrino momentum. We assign
$\etmiss$ to the transverse components of the neutrino momentum and compute the longitudinal
component from a $W$~boson mass constraint~\cite{Kane:1989vv}.
The longitudinal momentum of the neutrino $p_{\nu L}$ is formally expressed as
\begin{eqnarray}
p_{\nu L}  =\frac{1}{2\, p_{e T}^2}
 \left( {A\, p_{e L} \pm E_e \sqrt{A^2  - 4\,{p}^{\,2}_{e T}\etmiss^{\,2}}} \right),
\end{eqnarray}
where $A = M_W^2 + 2 \,\vec{p}_{e T} \cdot \,\vec{\etmiss} $.  
If $A^2 - 4 \, {p}^{\,2}_{e T}\,\etmiss^{\,2} > 0$, then 
there are two solutions and we pick the one with smaller $|p_{\nu L}|$. 
Otherwise the square root is complex and we pick the real part only.

%For the ambiguity to identify the jet from the top decay, we use the template method.

With the neutrino identified properly, we reconstruct the mass of the $B'$~quark as
\begin{equation}
m_{B'}^{\rm rec} = m(\vec{p}_{\nu} + \vec{p}_l + \vec{p}_{\rm jet\, 1}
               + \vec{p}_{\rm jet\, 2} + \vec{p}_{\rm jet\, 3})
\end{equation}

We then impose a window cut on the invariant mass difference between the reconstructed
invariant mass and the theory $B'$ mass under consideration,
\begin{eqnarray}
\left|m_{B'}^{\rm rec} - m_{B'}^{\rm theory}\right| < 100 \textrm{ GeV}\,. 
 \label{eqDQ:invmasscut}
\end{eqnarray}

Table~\ref{tabDQ:cs} shows the number of events passing each set of cuts, in units of fb.
~
\begin{table}[!h!tbp]
\caption{Cross sections for signal background processes at the $7$ TeV LHC passing
selection cuts.}
\label{tabDQ:cs}
\begin{center}
\renewcommand{\arraystretch}{1.4}
\begin{tabular}{lrrrrrrrr}
%\begin{tabular}{lcccccccc}
$\sigma$ [fb] & Signal  & $t +\rm{jets}$ & $t +W$ & $t\bar{t}$ & $t\bar{t}j$ & $WV$ & $W+\rm{jets}$ & total Bkg.  \\ 
no cuts       &  1062	&  18,877        & 2,861  & 22,200   & 7,900 	 & 10,007 & 2,457,400 & 2,519,245	  \\ 
basic cuts   &  507   	&  4,035      	&  808    & 5,491   & 772 	 & 1,692 & 92,521     & 105,319 \\ 
$+$~jet $p_T$ cuts & 346&  282     	&  163    & 3,117   & 297   	 & 324   & 27,645     & 31,828  \\
$+$~$H_T$ cuts &  295   &  100     	&  46     & 1,163   & 205 	 & 132   & 13,120     & 14,766 \\ 
$+$~$b$-tagging &  177  &  48           &  27     & 552     & 90         & 34 	 & 294        & 1,045 \\
$+$~mass window &  156   &  18           &  10     & 151     & 30         & 12    & 87         & 308
\end{tabular}
\end{center}
\end{table}

About half of the signal events pass the basic selection cuts. Only a third of the signal
events pass the final selection cuts including $b$-tagging, but the background is reduced
by a factor of 342. In particular the $b$-tagging cut reduces the $W$+jets background 
significantly. The mass window cut leaves a signal:background of 1:2 with sufficient
events remaining to be able to discover or rule out a $B'$ at this mass. Even for a
$B'$~mass of 1~TeV there are still 12~fb events remaining after the mass window cut, with
a background that is reduced by a factor of two, hence LHC searches should be sensitive
to this mass range with the data already recorded in 2011.

\begin{figure}[!h!tbp]
\begin{center}
	\includegraphics[scale=0.68]{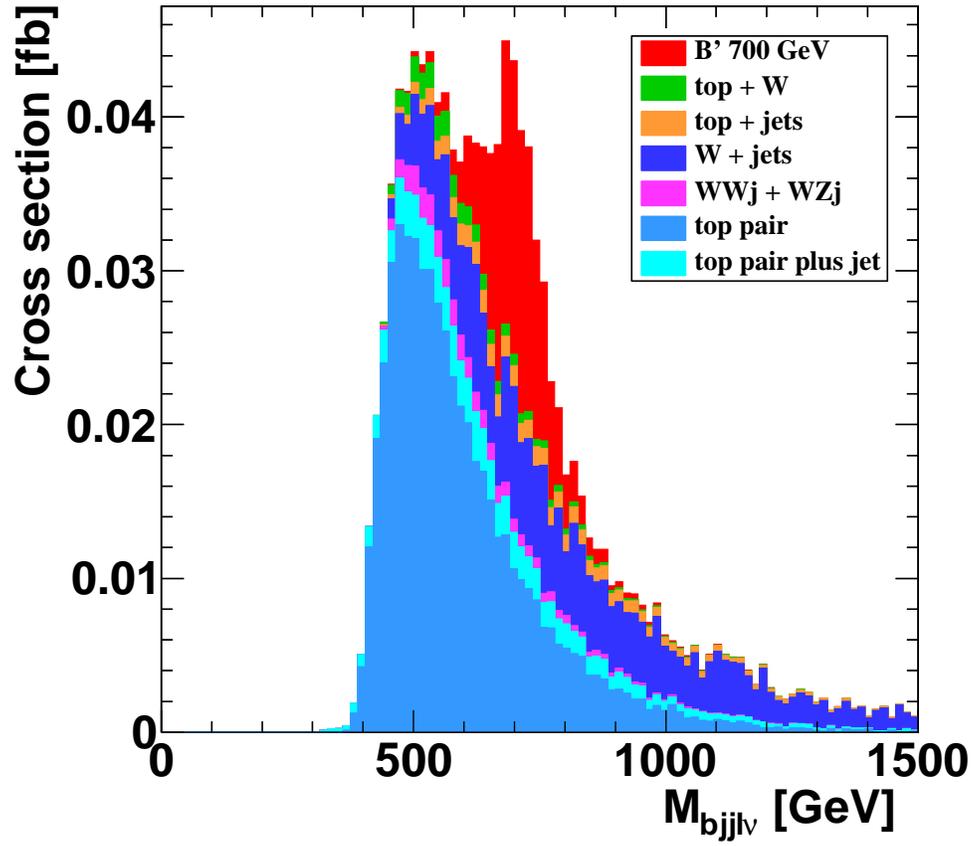}
	\caption{Reconstructed invariant mass for signal and backgrounds in the $B'$ 
analysis. The signal and backgrounds are normalized to their production cross sections.}
	\label{fig:sysmt}
\end{center}
\end{figure}

Figure~\ref{fig:sysmt} shows the reconstructed invariant mass. The signal peak is
clearly visible above the smoothly falling background.
~

\begin{figure}[!h!tbp]
\begin{center}
	\includegraphics[scale=0.6]{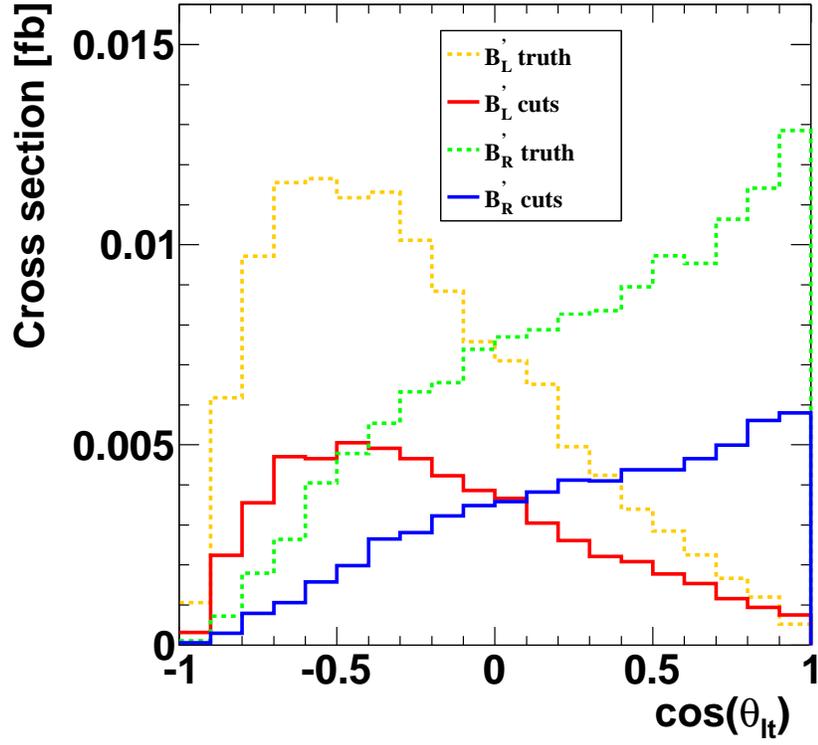}
	\caption{Angular correlation $\cos\theta$ between the final state lepton and the 
top moving direction for two different $B'$ models. }
	\label{figWP:sysmt}
\end{center}
\end{figure}

If a $B'$ is found, then it is possible to determine if it has left-handed or right-handed
couplings by looking at the $W$~boson helicity from the top quark decay. 
Figure~\ref{figWP:sysmt} shows the $cos \theta_{lt}$ distribution, where $\theta_{lt}$ is
the angle between the lepton in the top quark rest frame and the top quark moving direction
in the c.m. frame. At the parton (truth) level before any selection cuts, this results in
the familiar SM-like distribution for left-handed $B'$. The right-handed $B'$ distribution
is quite different, and the clear distinction remains even after selection cuts.
~

%
% Conclusions
%
\section{Conclusions}
We have presented a model for fourth generation $B'$~quarks and their single production
at the LHC with an effective Lagrangian that results in a $B'gb$ coupling. We have presented
the $B'$ decay branching ratios and production cross section at the LHC at several c.m. energies.
A phenomenological analysis shows that the LHC is sensitive to $B'$~quark production and decay
to a single top quark and $W$~boson. The experimental reach should be better than 700~GeV
already at the 7~TeV LHC, which makes this a very promising search channel. Moreover, once
a $B'$ quark is found, spin correlation in the $Wt$~final state can be used to determine
whether it is left-handed or right-handed.

\begin{acknowledgments}
The work of J.N. and R.S. was supported in part by the U.S. National Science Foundation under Grant No. PHY-0952729. D.W. is supported in part by a grant from the US National Science Foundation, grant NSF-PHY-0705682, the LHC Theory Initiative. The work of J.H.Y. was supported in part by the U.S. National Science Foundation under Grant No. PHY-0855561. 
\end{acknowledgments}

\clearpage    

\section*{Appendix}

\subsection*{A. Generalized B' Couplings with the SM and the Associated Branching Fractions }

Assuming the electroweak anomalous magnetic moment is suppressed compared to the 
colored anomalous magnetic moment, the Lagrangian for the electroweak couplings of the 
$B'$~quark is
\begin{eqnarray}
	{\mathcal L} &=& \frac{g_2}{\sqrt{2}} W^-_{\mu} \bar{t} \gamma^{\mu} (f_L P_L + f_R P_R) B'  + \frac{g_2}{2c_W} \,Z_{\mu}\, \bar{b} \gamma^{\mu} (F_L P_L + F_R P_R) B'  \nonumber \\
	&+& \frac{m_b}{v}\, h \,\bar{b}\, (y_L P_L + y_R P_R) B' + h.c.\, ,
\end{eqnarray}
where $\frac{m_b}{v} = \frac{g_2}{2}\frac{m_b}{m_W}$.  As a reminder, $g_2$ is the $SU(2)_L$ 
coupling and $m_b$ is the bottom quark mass.  Here $f_{L,R}$, $F_{L,R}$ and $y_{L,R}$ 
parametrizes the chirality of the $B'$~coupling with the different SM bosons.  
$P_{L,R}$ are the traditional 
projection operators.  It is straightforward to compute 
the partial decay widths of the $B'$~quark as
\begin{eqnarray}
	\Gamma (B' \to b\, Z) &=&   \frac{g_2^2}{128\pi c_W^2} \frac{M_{B'}^3}{M_Z^2} (F_L^2 + F_R^2) (1 - x_Z^2 )^2 ( 1 +  2 x_Z^2 ), ,\\
	\Gamma (B' \to t \,W^- ) &=& \frac{g_2^2}{64\pi} \frac{M_{B'}^3}{M_W^2} (f_L^2 + f_R^2) (1  -  x_t^2)^3 + {\mathcal O}(x_W^2) \, ,\\
	\Gamma (B' \to b \,h ) &=& \frac{g^2}{128\pi} \frac{M_{B'}^3}{M_W^2} (y_L^2 + y_R^2) (1 - x_h^2)^2 \, ,\\
	\Gamma (B' \to b \,g ) &=& \frac{g_s^2}{12\pi} M_{B'}  (\kappa_L^2 + \kappa_R^2) \, .
\end{eqnarray}
Here we define $x_Z = M_Z/M_{B'}$,  $x_W = M_W/M_{B'}$, $x_h = M_h/M_{B'}$ and $x_t = M_t/M_{B'}$.  As a reminder, $g_s$ is the strong coupling constant.  
Note that the general decay width $F\to f \,h$   is
\begin{eqnarray}
	\Gamma (F \to f \,h) &=&  \frac{g^2}{32\pi} M_F \lambda^{1/2}\left(x_h^2, x_f^2 \right)
	\left[ (y_L^2 + y_R^2) (1  + x_f^2 - x_h^2) + 4 x_f \textrm{Re}(y_L y^*_R)\right] \, ,
\end{eqnarray}
where $\lambda$ is given by
\begin{equation}
\lambda \left(x_h^2, x_f^2 \right) = 1 + x_h^4 + x_f^4 - 2 x_h^2 - 2 x_f^2 -2 x_h^2x_f^2 \, .
\end{equation}

%%%%%%%%%%%%%%%%%%%%%%%%%%%%%%%%%%%%%%%%%%%%%%%%%%%%%%%%%%%%%%%%%%%%%%%%%%%%%%%%%%%%%%%%%%%%%%%%%%%
%   References
%%%%%%%%%%%%%%%%%%%%%%%%%%%%%%%%%%%%%%%%%%%%%%%%%%%%%%%%%%%%%%%%%%%%%%%%%%%%%%%%%%%%%%%%%%%%%%%%%%%


\begin{thebibliography}{99}


\bibitem{atlashiggsresults}
   ATLAS Collaboration, ATLAS-CONF-2011-163.
  CERN Higgs Search Update Seminar,  ``Status of Standard Model Higgs Search in ATLAS, " July 4, 2012.

\bibitem{CMShiggsresults}
    CMS Collaboration, CMS-PAS-HIG-11-032,
  CERN Higgs Search Update Seminar,  ``Status of the CMS SM Higgs Search, " July 4, 2012.

\bibitem{technicolor}
  S.~Weinberg, 
  %``Implications Of Dynamical Symmetry Breaking,''
  Phys.\ Rev.\ D {\bf 13}, 974 (1976);
  %%CITATION = PHRVA,D13,974;%%
  L.~Susskind,
  %``Dynamics Of Spontaneous Symmetry Breaking In The Weinberg-Salam Theory,''
  Phys.\ Rev.\ D {\bf 20}, 2619 (1979).
  %%CITATION = PHRVA,D20,2619;%%

\bibitem{topcolor}
  C.~T.~Hill,
  %``Quark And Lepton Masses From Renormalization Group Fixed Points,''
  Phys.\ Rev.\ D {\bf 24}, 691 (1981); 
  %%CITATION = PHRVA,D24,691;%%
  C.~T.~Hill,
  %``Topcolor: Top Quark Condensation In A Gauge Extension Of The Standard
  %Model,''
  Phys.\ Lett.\ B {\bf 266}, 419 (1991);
  %%CITATION = PHLTA,B266,419;%%
  R.~S.~Chivukula, B.~A.~Dobrescu, H.~Georgi and C.~T.~Hill,
  %``Top quark seesaw theory of electroweak symmetry breaking,''
  Phys.\ Rev.\ D {\bf 59}, 075003 (1999).
  %%CITATION = HEP-PH 9809470;%%
  
%\cite{Dobrescu:1997nm}
\bibitem{topseesaw}
  B.~A.~Dobrescu and C.~T.~Hill,
  %``Electroweak symmetry breaking via top condensation seesaw,''
  Phys.\ Rev.\ Lett.\  {\bf 81}, 2634 (1998).
  %%CITATION = HEP-PH 9712319;%%  
  
\bibitem{tcreview}
  C.~T.~Hill and E.~H.~Simmons,
  %``Strong dynamics and electroweak symmetry breaking,''
  Phys.\ Rept.\  {\bf 381}, 235 (2003)
  [Erratum-ibid.\  {\bf 390}, 553 (2004)]
  and references therein.
  %%CITATION = HEP-PH 0203079;%%
  
\bibitem{Kaplan:1983sm}
  D.~B.~Kaplan, H.~Georgi and S.~Dimopoulos,
  %``Composite Higgs Scalars,''
  Phys.\ Lett.\ B {\bf 136}, 187 (1984).
  %%CITATION = PHLTA,B136,187;%%

\bibitem{LH}  
  N.~Arkani-Hamed, A.~G.~Cohen and H.~Georgi,
  %``Electroweak symmetry breaking from dimensional deconstruction,''
  Phys.\ Lett.\ B {\bf 513}, 232 (2001);
  %%CITATION = HEP-PH 0105239;%%
   N.~Arkani-Hamed, A.~G.~Cohen, E.~Katz and A.~E.~Nelson,
%``The littlest Higgs,''
  JHEP {\bf 0207}, 034 (2002).  

\bibitem{Arkani-Hamed:1998rs}
  N.\ Arkani-Hamed, S.\ Dimopoulos and G.~R.\ Dvali,
  %``The hierarchy problem and new dimensions at a millimeter,''
  Phys.\ Lett.\ B {\bf 429}, 263 (1998).
%  [arXiv:hep-ph/9803315].
  %%CITATION = HEP-PH 9803315;%%

\bibitem{Randall:1999ee}
  L.~Randall and R.~Sundrum,
  %``A large mass hierarchy from a small extra dimension,''
  Phys.\ Rev.\ Lett.\  {\bf 83}, 3370 (1999).
%  [arXiv:hep-ph/9905221].
  %%CITATION = HEP-PH 9905221;%%

%\cite{Chatrchyan:2011vp}
\bibitem{Chatrchyan:2011vp} 
  CMS Collaboration,
  %``Measurement of the t-channel single top quark production cross section in pp collisions at sqrt(s) = 7 TeV,''
  Phys.\ Rev.\ Lett.\  {\bf 107}, 091802 (2011).
  %[arXiv:1106.3052 [hep-ex]].
  %%CITATION = ARXIV:1106.3052;%%

%\cite{Aad:2012ux}
\bibitem{Aad:2012ux} 
  %G.~Aad {\it et al.}  [
  ATLAS Collaboration,
  %``Measurement of the t-channel single top-quark production cross section in pp collisions at sqrt(s) = 7 TeV with the ATLAS detector,''
  arXiv:1205.3130 [hep-ex].
  %%CITATION = ARXIV:1205.3130;%%

%\cite{ATLAS:2012dj}
\bibitem{ATLAS:2012dj} 
  ATLAS Collaboration,
  %``Evidence for the associated production of a W boson and a top quark in ATLAS at sqrt(s) = 7 TeV,''
  arXiv:1205.5764 [hep-ex].
  %%CITATION = ARXIV:1205.5764;%%

% single top at NLO and with NNLO corrections
%\cite{Kidonakis:2011wy}
\bibitem{Kidonakis:2011wy} 
  N.~Kidonakis,
  %``Next-to-next-to-leading-order collinear and soft gluon corrections for t-channel single top quark production,''
  Phys.\ Rev.\ D {\bf 83}, 091503 (2011);
  %[arXiv:1103.2792 [hep-ph]];
  %%CITATION = ARXIV:1103.2792;%%
%\cite{Kidonakis:2010tc}
%\bibitem{Kidonakis:2010tc} 
  N.~Kidonakis,
  %``NNLL resummation for s-channel single top quark production,''
  Phys.\ Rev.\ D {\bf 81}, 054028 (2010);
  %[arXiv:1001.5034 [hep-ph]];
  %%CITATION = ARXIV:1001.5034;%%
%\cite{Kidonakis:2010ux}
%\bibitem{Kidonakis:2010ux} 
  N.~Kidonakis,
  %``Two-loop soft anomalous dimensions for single top quark associated production with a W- or H-,''
  Phys.\ Rev.\ D {\bf 82}, 054018 (2010).
  %[arXiv:1005.4451 [hep-ph]].
  %%CITATION = ARXIV:1005.4451;%%

% ATLAS search for sm-like b'
%\cite{ATLAS:2012aw}
\bibitem{ATLAS:2012aw} 
  ATLAS Collaboration,
  %``Search for down-type fourth generation quarks with the ATLAS detector in events with one lepton and high transverse momentum hadronically decaying W bosons in sqrt(s) = 7 TeV pp collisions,''
  arXiv:1202.6540 [hep-ex].
  %%CITATION = ARXIV:1202.6540;%%

% CMS search for 4th generation
%\cite{Chatrchyan:2012yea}
\bibitem{Chatrchyan:2012yea} 
  CMS Collaboration,
  %``Search for heavy bottom-like quarks in 4.9 inverse femtobarns of pp collisions at sqrt(s) = 7 TeV,''
  JHEP {\bf 1205}, 123 (2012).
  %[arXiv:1204.1088 [hep-ex]].
  %%CITATION = ARXIV:1204.1088;%%

% CDF b' search
%\cite{Aaltonen:2011vr}
\bibitem{Aaltonen:2011vr} 
  CDF Collaboration,
  %``Search for heavy bottom-like quarks decaying to an electron or muon and jets in $p\bar{p}$ collisions at $\sqrt{s}=1.96$ TeV,''
  Phys.\ Rev.\ Lett.\  {\bf 106}, 141803 (2011).
  %[arXiv:1101.5728 [hep-ex]].
  %%CITATION = ARXIV:1101.5728;%%

% D0 vector quark search
%\cite{Abazov:2010ku}
\bibitem{Abazov:2010ku} 
  %V.~M.~Abazov {\it et al.}  [
  D0 Collaboration,
  %``Search for single vector-like quarks in $p\bar{p}$ collisions at sqrt(s) = 1.96 TeV,''
  Phys.\ Rev.\ Lett.\  {\bf 106}, 081801 (2011).
  %[arXiv:1010.1466 [hep-ex]].
  %%CITATION = ARXIV:1010.1466;%%

% ATLAS b' to Zb search
%\cite{ATLAS:2012ak}
\bibitem{ATLAS:2012ak} 
  ATLAS Collaboration,
  %``Search for pair production of a new quark that decays to a Z boson and a bottom quark with the ATLAS detector,''
  %Submitted to: Phys.Rev.Lett.
  [arXiv:1204.1265 [hep-ex]].
  %%CITATION = ARXIV:1204.1265;%%

%\cite{Aad:2011yn}
\bibitem{Aad:2011yn} 
  ATLAS Collaboration,
  %``Search for heavy vector-like quarks coupling to light quarks in proton-proton collisions at sqrt(s) = 7 TeV with the ATLAS detector,''
  Phys.\ Lett.\ B {\bf 712}, 22 (2012).
  %[arXiv:1112.5755 [hep-ex]].
  %%CITATION = ARXIV:1112.5755;%%

% Chanowitz
%\cite{Chanowitz:1978uj}
\bibitem{Chanowitz:1978uj} 
  M.~S.~Chanowitz, M.~A.~Furman and I.~Hinchliffe,
  %``Weak Interactions of Ultraheavy Fermions,''
  Phys.\ Lett.\ B {\bf 78}, 285 (1978).
  %%CITATION = PHLTA,B78,285;%%

%\cite{Chanowitz:1978mv}
\bibitem{Chanowitz:1978mv} 
  M.~S.~Chanowitz, M.~A.~Furman and I.~Hinchliffe,
  %``Weak Interactions of Ultraheavy Fermions. 2.,''
  Nucl.\ Phys.\ B {\bf 153}, 402 (1979).
  %%CITATION = NUPHA,B153,402;%%

%\cite{Peskin:1991sw}
\bibitem{Peskin:1991sw} 
  M.~E.~Peskin and T.~Takeuchi,
  %``Estimation of oblique electroweak corrections,''
  Phys.\ Rev.\ D {\bf 46}, 381 (1992).
  %%CITATION = PHRVA,D46,381;%%

\bibitem{Nakamura:2010zzi} 
  K.~Nakamura {\it et al.}  [Particle Data Group Collaboration],
  %``Review of particle physics,''
  J.\ Phys.\ G G {\bf 37}, 075021 (2010).
  %%CITATION = JPHGB,G37,075021;%%

%\cite{Agashe:2006at}
\bibitem{Agashe:2006at} 
  K.~Agashe, R.~Contino, L.~Da Rold and A.~Pomarol,
  %``A Custodial symmetry for Zb anti-b,''
  Phys.\ Lett.\ B {\bf 641}, 62 (2006).
  %[hep-ph/0605341].
  %%CITATION = HEP-PH/0605341;%%

%\cite{SekharChivukula:2009if}
\bibitem{SekharChivukula:2009if}
  R.~Sekhar Chivukula, S.~Di Chiara, R.~Foadi and E.~H.~Simmons,
  %``The Limits of Custodial Symmetry,''
  Phys.\ Rev.\ D {\bf 80} (2009) 095001
   [Erratum-ibid.\ D {\bf 81} (2010) 059902].
  %[arXiv:0908.1079 [hep-ph]].
  %%CITATION = ARXIV:0908.1079;%%

%\cite{Baur:1987ga}
\bibitem{Baur:1987ga} 
  U.~Baur, I.~Hinchliffe and D.~Zeppenfeld,
  %``Excited Quark Production at Hadron Colliders,''
  Int.\ J.\ Mod.\ Phys.\ A {\bf 2}, 1285 (1987).
  %%CITATION = IMPAE,A2,1285;%%

%\cite{Alwall:2011uj}
\bibitem{Alwall:2011uj} 
  J.~Alwall, M.~Herquet, F.~Maltoni, O.~Mattelaer and T.~Stelzer,
  %``MadGraph 5 : Going Beyond,''
  JHEP {\bf 1106}, 128 (2011). We use MadGraph~5 version 1.3.
  %[arXiv:1106.0522 [hep-ph]].
  %%CITATION = ARXIV:1106.0522;%%

%\cite{Pumplin:2002vw}
\bibitem{Pumplin:2002vw} 
  J.~Pumplin, D.~R.~Stump, J.~Huston, H.~L.~Lai, P.~M.~Nadolsky and W.~K.~Tung,
  %``New generation of parton distributions with uncertainties from global QCD analysis,''
  JHEP {\bf 0207}, 012 (2002).
  %[hep-ph/0201195].
  %%CITATION = HEP-PH/0201195;%%

%\cite{Cacciari:2011ma}
\bibitem{Cacciari:2011ma} 
  M.~Cacciari, G.~P.~Salam and G.~Soyez,
  %``FastJet user manual,''
  Eur.\ Phys.\ J.\ C {\bf 72}, 1896 (2012).
  %[arXiv:1111.6097 [hep-ph]].
  %%CITATION = ARXIV:1111.6097;%%

%\cite{Aad:2008zzm}
\bibitem{Aad:2008zzm} 
  %G.~Aad {\it et al.}  [
  ATLAS Collaboration,
  %``The ATLAS Experiment at the CERN Large Hadron Collider,''
  JINST {\bf 3}, S08003 (2008).
  %%CITATION = JINST,3,S08003;%%
  %arXiv:0901.0512 [hep-ex].

%%\cite{Bayatian:2006zz}
%\bibitem{Bayatian:2006zz} 
%  G.~L.~Bayatian {\it et al.}  [CMS Collaboration],
%  %``CMS physics: Technical design report,''
%  CERN-LHCC-2006-001.
%  %%CITATION = CERN-LHCC-2006-001;%%
%\cite{Ball:2007zza}
\bibitem{Ball:2007zza} 
  %G.~L.~Bayatian {\it et al.}  [
  CMS Collaboration,
  %``CMS technical design report, volume II: Physics performance,''
  J.\ Phys.\ G G {\bf 34}, 995 (2007).
  %%CITATION = JPHGB,G34,995;%%

%\cite{Kane:1989vv}
\bibitem{Kane:1989vv} 
  G.~L.~Kane and C.~P.~Yuan,
  %``How to study longitudinal W's in the TeV region,''
  Phys.\ Rev.\ D {\bf 40}, 2231 (1989).
  %%CITATION = PHRVA,D40,2231;%%

\end{thebibliography}
\end{document}